\documentstyle{article}
\newtheorem{result}{Result}
\newtheorem{theorem}{Theorem}
\newtheorem{lemmas}{Lemma}
\newtheorem{conj}{Conjecture}
\newcommand{\myl}[1]{\label{#1}\eeq}
\newcommand{\myll}[1]{\label{#1}}
\newcommand{\myc}[1]{\cite{#1}}
\newcommand{\myref}[1]{(\ref{#1})}
\newcommand{\mysref}[1]{Section\ \ref{#1}}

\newcommand{\beq}{\begin{equation}}
\newcommand{\eeq}{\end{equation}}
\newcommand{\beqa}{\beq\begin{array}{l}}

\newcommand{\mR}{${\bf R^4}$}

\newcommand{\mRn}{${\bf R^n}$}

\newcommand{\mRx}{${\bf R^4_\Theta}$}
\newcommand{\Rx}{\bf R^4_\Theta}
\newcommand{\msx}{$\bf\Sigma^7_\Theta$}
\newcommand{\sx}{\bf\Sigma^7_\Theta}
\newcommand{\xt}{\times_\Theta}
\newcommand{\dM}{${\cal D}$}

\newcommand{\bdM}{${\bar{\cal D}}$}

\newcommand{\DS} {differentiable structure}
\newcommand{\DSs} {differentiable structures}

\newcommand{\state}[2]{\begin{center}\vspace{\baselineskip}
\parbox{.85\textwidth}{{\bf #1:}\ {\em
#2}}\vspace{\baselineskip}\end{center}}
\begin{document}

\begin{titlepage}
\begin{center}
{\Large \bf  Exotic Smoothness and Physics}
\end{center}
\vspace{.5truein}
\begin{center}
{\large \bf Carl H. Brans}
\end{center}
\vspace{.5truein}
\begin{center}
Physics Department\\ Loyola University\\ New Orleans, LA 70118  \\
e-mail:BRANS @ MUSIC.LOYNO.EDU\par\bigskip
March 29, 1994
\end{center}
\begin{abstract}
The essential role played by \DSs\ in physics
 is reviewed in
light of recent mathematical discoveries that topologically
trivial space-time models, especially the simplest one, \mR,
possess a rich multiplicity of such structures, no two of which
are diffeomorphic to each other and thus to the standard one.
This means that physics has available to it a new
panoply of structures available for space-time models.  These can
be thought of as source of
new global, but not properly topological, features.  This
paper reviews some background differential topology
together with a discussion of the
role which a \DS\ necessarily plays in the statement of
any physical theory, recalling that
diffeomorphisms are at the heart of the principle of
general relativity. Some of the history of the discovery
 of exotic, i.e.,
non-standard, \DSs\ is reviewed.  Some new results suggesting
the
spatial localization of such exotic structures are described and
speculations are made on the
possible opportunities that such structures
present for the further development of physical theories.\par

\end{abstract} \par\bigskip
To be published in {\it Journal of Mathematical Physics}.
\end{titlepage}
\section{Introduction}\myll{secI}\par
Almost all modern physical theories are stated in terms of
fields over a smooth manifold, typically involving differential
equations for these fields.  Obviously the mechanism
underlying
the notion of differentiation in a space-time model is
indispensable for physics.
However, the essentially local nature of
differentiation has led us to assume that the mathematics
underlying calculus as applied to physics is trivial and unique.
In a certain sense this is true:  all smooth manifolds of the
same dimension are {\it locally} differentiably equivalent, and
thus physically indistinguishable in the local sense.
However, only fairly recently have  we learned that this
need not be true in the global sense.
Mathematicians have discovered that certain common, even
topologically trivial, manifolds may not be
{\it globally} smoothly equivalent (diffeomorphic)
even though they are {\it globally} topologically equivalent
(homeomorphic).  This is a strikingly counter-intuitive result
which presents new, non-topological, global possibilities for
space-time models. The fact
that the dimension four plays a central role in this story
only enhances the speculation that these results may be of
some physical significance.
\par
Briefly, until some twelve years ago it was known that any smooth
manifold that was
{\it topologically} equivalent to ${\bf R^n}$ for $n\ne 4$
 must necessarily be {\it differentiably} equivalent to it.
It is probably
safe to say most workers expected that the outstanding case of
$n=4$
would be resolved the same way when sufficient technology was
developed for the proof.  Thus, it was quite a shock when the
work of Freedman\myc{fr} and Donaldson\myc{D} showed that
there are an infinity of \DSs\ on topological \mR, no two of
which are equivalent, i.e., diffeomorphic, to each other.  The
surprising feature of this result is the fact that while each of
these manifolds is {\it locally} smoothly equivalent to any
other, this local equivalence cannot be continued to a global
one, even though the manifold model, \mR, is clearly
topologically trivial.   Among this infinity of smooth manifolds
homeomorphic to \mR\ is the standard one, with smoothness
inherited simply from the smooth
product structure ${\bf R^1\times
R^1\times R^1\times R^1}.$  This standard one we denote by the
same symbol as we have used for the topological model, \mR.  The
other smooth manifolds are called variously ``fake,''
``non-standard,'' or ``exotic.''  The last is the term used in
this paper and such a manifold is denoted by \mRx.
 \par
The fact that  \mRx's
arise only in the physically significant case of dimension four
makes the result even more intriguing to physicists.
Unfortunately, however, these proofs have been more
existential than constructive in
the sense of a practical coordinate patch
presentation. This lack makes progress
in the field difficult.  For example, no global expression can be
obtained for any non-scalar field, such as the metric.
However, certain existence, and non-existence,
 results can be obtained
and some of these will be reviewed in this paper.  \par
The intent of this paper is to provide a review of the
mathematical technology necessary to grasp the significance of
some of these results and to point out their possible
implications for physics.  In doing so, we draw extensively on
two earlier papers, \myc{br},\myc{blx}.
We begin with a review of some basic differential
topology  and discuss
its relevance to the expression of physical theories.  In
particular, we recall
 the notions of {\it topological} and
{\it smooth} manifolds and the important difference between them.
This difference
is often overlooked and the smoothness structure is simply
assumed to be some ``standard'' one naturally
associated with the
topological structure.  Along the way, the concept of a {\it
\DS\ } will be defined, involving first the notion of an {\it
atlas} and then the equivalence class under diffeomorphisms.
The distinction between ``different'' and ``non-diffeomorphic''
\DSs\ is so important that it will be exemplified
in terms of simple real models and then a toy model
in which an explicit ``exotic'' complex structure is explored
as an alternative to the standard one for two-dimensional vacuum
Maxwell's equations.
  \par
  Next, a brief review of the discovery of exotic smoothness
structures will be given, starting with the explicit case of
Milnor seven-spheres.
After this, the somewhat tortuous path leading to the discovery
of \mRx's is reviewed.  This story is much more involved and
calls upon a much wider variety of mathematical fields than that
of the Milnor spheres.
Finally, the development of tools enabling
the definition of a candidate for ``spatially localized'' exotic
structures will be described and discussed.  We close with some
conjectures and speculations.\par
\section{Manifolds: Topological and Smooth}\myll{m}\par
Our primitive construct for modelling space-time was for
a long time the
standard Euclidean manifold, \mR,  a special case of the
standard manifold,  ${\bf R^n}.$ This latter is defined as the
set of points,
${\bf p}$, each of which can be identified with an n-tuple of real
numbers, $\{p^\alpha\}$. The topology is induced by the
 usual product
topology of the real line.  What is important for our
discussion
is that this standard model implicitly imparts to each space-time
point an identifiable, objective, existence, independent of any
choice of coordinatization.  In fact, this topological notion is
{\it prior} to the existence of any coordinates\footnote{
This ``objective reality'' assigned to points in space-time
models is, of course, strongly reminiscent of the old
pre-relativity ether models.  However, we are not concerned
with such questions here.  Many people have raised this issue.
Some thoughts on it are contained in my article, ``Roles
of Space-Time Models,'' p27, in  {\it Quantum Theory and
Gravitation,} ed. A.R. Marlow, Academic Press, 1980.}.
  A thorough
understanding of this fact is absolutely necessary to follow
 the intricacies involved with the definition of
\DSs.  Historically, of course, Einstein and
others simply identified the list $\{p^\alpha\}$ as a list of
some physical coordinates thereby choosing what we will call the
standard \DS. By choosing this standard \mR,
we limit physics  to the
trivial
Euclidean topology, eliminating the rich and potentially very
useful set of physical possibilities inherent in non-trivial
topologies. Also, and more importantly for the purposes of this
paper, this choice also precludes investigation of
 the myriad possibilities
opened up by the recent discoveries in differential topology.
Consequently, we now proceed to review some of the basic tools
of this field.
\par
First, recall the notion of a {\it topological manifold}. A
topological space, $M$,
 is a topological manifold if it is locally
homeomorphic to \mRn.  Thus, $M$ must be covered by
a family of open sets, $\{U_a\}$,
each of which is homeomorphic to  \mRn.
Such a family is called a (topological) {\it atlas},
${\cal A}=\{U_a,x_a^i\},$ where the $x_a^i$ are the local
\mRn\ homeomorphisms.
Each $(U_a, x_a^i)$
in the atlas thus serves as topological coordinate
patch, or {\it chart},
 locally identifying points with ordered sets of $n$ real
numbers in a topological manner.  Unlike the case of smooth
manifolds defined below, note that no further
assumptions need be made concerning the transition functions
between two overlapping patches.  They will necessarily be
local homeomorphisms in standard \mRn.
Also note that all manifolds
 used in this paper are assumed to be Hausdorff.\par
To this point, no notion of differentiation has been introduced.
Thus, if $M$ is a topological manifold with elements (points)
{\bf p}, and $f$ is a real-valued map, $f:M\rightarrow{\bf R},$
there is not yet any structure available to do calculus with $f$.
Of course, the existence of the local charts identifying
 points, ${\bf p}$,
with n-tuples, $x_a^i,$
 of real numbers would seem to provide a clue.
For example, we can
re-express $f$ locally as
$f_a:x_a(U_a)\rightarrow {\bf R}$ given by
\beq
f_a(x^i)=f({\bf p}),\ \
{\rm where\ }x^i=x_a^i({\bf p}).\myl{m1}
We can then proceed to do calculus on the invariantly defined $f$
in terms of the well-known, standard, real variable calculus
applied to its coordinate patch form,
 $f_a.$
Using this representation, $f$ will be {\it smooth}, or
{\it differentiable} in the
 $C^\infty$ sense if and only if $f_a$
is $C^\infty$
in the standard real variable sense.  However,
 to be useful, this definition should be
{\it independent of the local chart}, $\{U_a,x_a^i\}$ used in
its
definition.  The condition for this consistency is contained in
the definition of a {\it smooth atlas of charts.}  Given
a topological manifold, $M$, a topological atlas
${\cal A}=\{U_a,x_a^i\}$ covering $M$
is said to be a {\it smooth atlas} if for every pair
for which  $W=U_a\cap U_b$ is not empty, the local
homeomorphism of \mRn\ defined by $x_a\cdot x_b^{-1}$  is
smooth
in the standard sense in \mRn.  This forms the basis for
differential topology and
 seems to contain the minimal
structure needed to do physics expressed in differential
equations. \par
Clearly any given $M$ could conceivably support many different
smooth atlases.  Two such, say ${\cal A}$ and ${\cal A'},$ are
said to be compatible if their union is
also a smooth atlas.  This
leads naturally to a set-theoretic type of ordering and the
notion of a maximal atlas.  Using this we now define a central
notion in differential topology.  A {\it \DS}, \dM$(M)$,
is a maximal smooth atlas on $M$, and defines $M$ as a {\it
 smooth, or differentiable, manifold.}
  Clearly any atlas in a
maximal one defines the \DS, so the maximalization procedure is
often not explicitly carried out. It  turns out that
most topological
manifolds can support more than one \dM,
so there are many smooth
manifolds over any given topological one.
A natural question to ask, and certainly an important one
for physics, is whether or not two different smooth manifolds
with the same topology are really equivalent in an appropriate
differentiable sense.\par
This question is answered in terms of the notion of a {\it
diffeomorphism.}  First, a
 map, $f:M\rightarrow M'$ of
one
smooth manifold into another is said to be {\it differentiable}
if
it is smooth in the standard \mRn\ sense when expressed in the
local smooth charts.  $f$ is a {\it diffeomorphism} if it is a
differentiable homeomorphism.
{}From the viewpoint of differential topology two smooth manifolds
are equivalent if they are diffeomorphic.  The same is true for
physics, since a diffeomorphism is the mathematical embodiment of
the notion of a global re-coordinatization of the manifold.
Thus, the principle of general relativity demands that the
physics of two different, but
diffeomorphic, manifolds be regarded as identical.
However, is the \DS\ determined by the topology? That is,
\par
{\bf Fundamental Question:\it Can two homeomorphic manifolds
support truly different, i.e., non-diffeomorphic, \DSs?}\par
If so, these manifolds describe space-time models which while
identical both in the global topological and local smooth senses
are {\it not} physically equivalent.\par
\section{Simple Examples and a Metaphor}\myll{exm}\par
This question opens the door to new possibilities, other
than topological, for global structures and
their effects to occur in
physical space-time models.  However, the difference between
  {\it different} \DSs, and {\it non-diffeomorphic} ones, is a
somewhat elusive concept, so in this section some simple examples
will be considered.
Also, a toy model involving the more easily managed complex
structures on the plane will be explored.  This allows us
to look explicitly at the distinction between {\it different}
complex structures and those which produce manifolds which are
not {\it biholomorphic} to each other.\par
First, consider the problem of establishing a \DS\ on the simple
topological manifold, ${\bf R^1}.$  Recall the
distinctive notation to identify the ``objectively existing''
points of this space, ${\bf R^1}=\{{\bf p}\}$ where, in this case
each point, {\bf p}, is simply one real number, $p$.
  A natural
maximal atlas and thus a \DS, \dM, is generated simply by the
one global chart with coordinate, $x({\bf p})=p.$  This is the
standard \DS\ for ${\bf R^1}$.
In this \DS, a function of {\bf p} is smooth if and only if it is
a smooth function of $p$ in the usual sense.
 However, there are many other possible \DSs. For example, one
such is generated by the same global chart, but with coordinate
$y({\bf p})=p^{1\over 3}.$
Clearly this leads to a {\it different} \DS, say
\bdM, since the combination $y\cdot x^{-1}:p\rightarrow
p^{1\over 3},$
which is not smooth.  So, the topological manifold, ${\bf R^1}$
has a least two {\it different} \DSs\ on it, \dM\ and \bdM.
However, it is easy to show that these different structures would
not lead to different physics (or smooth mathematics
for that matter) since they are actually {\it diffeomorphic.}
This result can be explicitly demonstrated by
defining the homeomorphism, $f$,
 of ${\bf R^1}$ onto itself by $f(p)=p^3.$  This is a
diffeomorphism since its expression in the two charts is $y\cdot
f\cdot x^{-1}:p\rightarrow
(p^3)^{1\over 3}=p$ is clearly smooth.
Thus, \dM\ and \bdM, while being {\it different} \DSs,
 are in fact {\it diffeomorphic}.  Actually, it is well
known
and fairly easy to prove that ${\bf R^1}$ can support only one
\DS\ up to diffeomorphisms.  Thus, no new
 {\it global} physics on
one-dimensional models can be encountered without changing the
 topology itself.
At first glance, the idea that $f(p)=p^3$ should be a
diffeomorphism may be somewhat troubling.  However, the issue is
that the original numbers $p$ do not of themselves define a
basis for differentiation, {\it unless an explicit assumption is
made}.  This assumption is that the \dM\ is the \DS\ to be used.
However, there is nothing any more natural to this assumption
than there is to the assumption that our space-time geometry
should be flat.
  \par
{}From this simple example it is quite tempting to conjecture that
while topologically trivial spaces such as \mRn\ can support an
infinity of {\it different} \DSs, none of these will lead to any
new physics since each will be diffeomorphic to the standard one.
If this were true, then there would likely be no new physics from
{\it differential} topology since all global structures would
have to be relegated to the topology.
As discussed in the introduction this is
in fact the case for all $n\ne 4.$ However, the surprising
discovery that this conjecture is not true precisely for the
physically significant dimension $n=4,$ means that potential new
global physical tools are available, apart from topological ones.
Unfortunately, the absence of any effective coordinate patch
presentation of any exotic smoothness makes the physical
exploration difficult.  However, we do have access to a
manageable model which can serve illustrate
the relationships that are important, but difficult
to make explicit, in the smooth
case.  This is provided by the notion of complex structures and
related holomorphisms.
The  problem  of  {\em complex}  structures  on ${\bf R}^2$ is
relatively easy, compared to that of the {\em differentiable}
 structures on ${\bf R}^4$ , so we can explicitly  explore  the
relationship  between the
mathematical structure and its physical implications.
\par
     Consider then the two-dimensional physics of vacuum
electrostatic  vector  fields  ${\bf
E}(x,y)$
described by component functions $E_x(x,y)$ and $E_y(x,y)$.  The Maxwell
vacuum electrostatic equations are
just the Cauchy conditions for the real
and (negative) imaginary parts of  an  analytic  function  of  the
complex variable $z\equiv  x  +  i\  y$.
The most general solution
to the Maxwell equations          in this formalism
 can be obtained simply from the complex
equation,
\beq
{\rm\bf Vacuum\ 2-D\ Maxwell}\Leftrightarrow
E_x - i\ E_y=F(z),\myl{cs3}
where $F(z)$ is an arbitrary analytic function.
\par
These facts are  well-known and
apart from  a  few  illustrative  boundary
value problems, do not seem to lead to any significant
physical
consequences   or   further   insights,   probably   because   the
introduction of a complex structure on the space model is possible
only for two-dimensional problems. However, the fact that
all structures can be explicitly described make this an excellent
example to illustrate the concepts in the real, smooth case.
\par
Complex structures are defined in a manner analogous to that of
smooth structures, with complex analyticity replacing
differentiability, and biholomorphisms replacing diffeomorphisms.
Thus  a  complex  structure, $CS$,  on a
two-dimensional
manifold, $M$, is defined by covering $M$ with  an atlas of
charts,
$U_a$ , together  with  maps,  $z_a$   taking $U_a$ (smoothly
and
invertibly) onto open balls in ${\bf R}^2$\ identified with the complex
plane ${\bf C}$ in the ``standard'' way, so that
 the local coordinates can be expressed
$z_a:U_a\rightarrow x_a({\bf p})+iy_a({\bf p})
\varepsilon {\bf C},$
for ${\bf p}\varepsilon U_a\in M.$
Furthermore, where defined, $z_a\circ z^{-1}_b$ must be analytic
in ${\bf C}$ in the usual complex sense.  The charts, $U_a$, are
sometimes
called ``coordinate patches'' and, for
 ${\bf p}\varepsilon U_a \subset
M$,
the value  $z_a({\bf p})\varepsilon {\bf C}$ is the
``coordinate of {\bf p}
relative to the patch $U_a$.''  A complex valued function,
$F:V\rightarrow{\bf C}$, for some neighborhood $V\subset M$, is
``analytic'', or ``holomorphic'', if it is complex analytic when
expressed
in the local coordinates, $z_a$, over the $U_a$ covering $V$,
  that is,
$F\circ z^{-1}_a$ is analytic (where defined) in the usual sense
on
${\bf C}$.
\par
Two such structures on a given $M$, say
$CS^\prime$ given by $\{U^\prime_a,z^\prime_a\}$ and $CS$ given
by $\{U_a,z_a\}$ are {\em analytically
equivalent (biholomorphic)} if and only if
there exists a homeomorphism, $F$, of $M$ onto itself such that
when $F$ is expressed in terms of the local charts it is a local
biholomorphism in the standard sense.  Thus,
$z_a\circ F\circ  z^{\prime-1}_b$
 and $z^\prime_a\circ F^{-1}\circ  z^{-1}_b$
must both be holomorphic where defined.  Note that it is not
necessary that the $CS$ coordinates themselves
be analytic in terms of $CS^\prime$,
but only when combined with a homeomorphism.
This is an important distinction. To clarify this point, consider
$U_1=U^\prime_1={\bf R^2}$, with $z_1(x,y)=x+i\ y$, and
$z^\prime_1(x,y)=x-i\ y$.  Then clearly the primed
coordinate is not
analytic in terms of the unprimed one.  However, these are
equivalent complex structures since the homeomorphism,
$F(x,y)=(x,-y)$ satisfies the above condition for equivalence.  That is,
\beq
z_1\circ F\circ z^{\prime-1}_1:z=x+iy\rightarrow
(x,-y)\rightarrow
(x,y)\rightarrow x+iy=z.\myl{cs6}
Thus, to the extent that the complex structure determines the
physics,
the physical content of $CS$ is identical to that of
$CS^\prime$.
For example, \myref{cs3}  could be expressed in {\it either}
$CS$
or $CS^\prime$ with the same physical content since the
homeomorphism $F$ takes us from one expression of a physical
electrostatic field to another.   In this complex
model, the invariance of the physical theory under
the transition from $CS$ to $CS^\prime$ is precisely
 analogous to the diffeomorphism invariance
 required in the real \DS\ case as formulated in
the principle of general relativity.
  It is easy to think
that there is something preferred about the standard complex
structure so that notions of analyticity must be settled in
terms of $z=x+iy$\ only.  However, this is clearly not so,
but the prejudice that there should be is reminiscent of that in
the real \DS\ case.
\par
Of course, this model would be trivial for our purposes if the
standard $CS$ were unique up to biholomorphisms.
However, the standard complex structure  is  not  unique!
There is
precisely one other inequivalent one, that is, one
{\it not} related
by a biholomorphism.   One  presentation  of  this
second  structure, $CS_1$,  can   be   defined   by
\beq
(x,y)\rightarrow z_1=
{e^{-1/r}\over r}(x+i\ y)\varepsilon {\bf C}.\myl{cs5}
The key point here is that $\vert z_1\vert$ is bounded.
It is now easy to show that $CS_0$ is not equivalent to $CS_1$.  In fact,
if it were then there would exist a function,
${\bf F}(x,y)=(F_x(x,y),F_y(x,y))$,
of the plane onto itself such that
${e^{-1/F}\over F}(F_x(x,y)+i\ F_y(x,y))$
 would be a global analytic function
of $x+i\ y$ in the usual sense.  Clearly however this cannot be since
this function is non-constant, but bounded on the entire plane, violating
a well known property of global analytic functions for the
{\it standard} $CS_0$.
\par
Now we can state the physical implications of the choice of structure,
complex in this case: If the physical theory is expressed by the
statement that the $x$ and $y$ components of the electrostatic vacuum
two-dimensional field are real and (negative) imaginary parts of a
function  analytic relative to the chosen complex structure,
then $CS_0$ and $CS_1$ lead to different fields, with physically
measurable differences since non-constant electrostatic
fields in the $CS_0$ case cannot be bounded, whereas they
are in the $CS_1$ case.
 In other words, to the extent that \myref{cs3} is the statement
 of a physical theory, then
in principle, experiment could
distinguish $CS_0$ from $CS_1.$
  However, experiment cannot distinguish
$CS_0$ from $CS^\prime$ described earlier, since these are biholomorphic.
\par
Of course,  this discussion is intended to be illustrative
rather
than of any likely physical significance itself.  The description
of electrostatic field theory in terms of analyticity requirements
is
certainly not the basis of a general physical theory.  In fact, it could
be argued
that changing the complex structure results in a changed metric and
that the correct theory should include this metric.
However, this model does illustrate explicitly the difference
between merely {\it different} and {\it inequivalent} structures,
non-biholomorphic in this case, non-diffeomorphic in our smooth
case.  For this complex model, $CS_1$ plays the role of an
exotic \DS\ in the real case. While $CS_1$ was easily displayed
we do not have yet have this luxury for a \mRx.   However,
we now proceed to review the path to these strange structures.
\par
\section
{Some History of Exotic Differentiable Structures}\myll{exds}
\par
An early and fairly easily accessible
example of an  exotic differentiable
structure on a simple topological space was provided
in 1956 by Milnor\myc{miln}.  He was able to use an
extension of the Hopf fibering of spheres\myc{st} to construct an
{\it exotic seven-sphere}, $\sx.$
 Consider the $S^3$ bundles over $S^4$,
\beq
\begin{array}{lrl}
S^3 & \rightarrow & M^7 \\
&  p & \downarrow \\
& & S^4
\end{array}\myl{b1}
with $Spin(4)$ (the covering group of
the rotation group, $SO(4)$) acting on $S^3$, as bundle group.
In its original form, the Hopf fibering makes use of the fact
that the topological seven sphere can be expressed in terms of a
pair of quaternions, $S^7=\{(q_1,q_2):\vert q_1^2\vert +
\vert q_2^2\vert=1\}.$  The projection to $S^4$ is through the
quaternion projective map: $\{q_1:q_2\}\rightarrow S^4.$  The
natural kernel of this map is of course the set of unit
quaternions, $S^3=SU(2)\subset Spin(4).$  For a wider class of
bundles \myref{b1}
a classification  is provided by
 $\pi_3(Spin(4))\approx Z+Z,$
as discussed in \ \S 18 of    \myc{st}.
This construction can be described in terms of the normal form for $M^7$
in which the base $S^4$ is covered by two coordinate patches, say upper
and lower hemispheres.  The overlap is then ${\bf R^1}\times S^3$ which
has $S^3$ as a retract.  Thus, the bundle transition functions are
defined by their value on this subset, defining a map from $S^3$ into
$Spin(4)$ and thus generating an element of
$\pi_3(Spin(4)).$  The group
action of $Spin(4)$ on the fiber $S^3$ can  be conveniently described in
the well-known quaternion form,
$u\rightarrow u'=vu\overline w,$
where $u,v,w$ are all unit quaternions and $\overline w$ is quarternion
conjugate of $w$.  Thus $u\in S^3$ and $(v,w)\in SU(2)\times SU(2)\approx
Spin(4).$ Standard $S^7$ is obtained from the element of $\pi_3(Spin(4))$
generated by $(v,1)$, so that the group action reduces to one
$SU(2)\approx S^3$ and the bundle is in fact an $SU(2)$ principal one.
In fact, this is precisely the principle $SU(2)$ Yang-Mills bundle over
compactified space-time, $S^4.$  For more details on classifying sphere
bundles, see
\myc{st},\ \S 20 and, from the physics viewpoint,  \myc{traut}.\par
Milnor's breakthrough in 1956 involved his proving that $M^7$ for the
transition function map, an element of $\pi_3(Spin(4)),$ given by
\beq
  u\rightarrow (u^h,\overline u^j)\in Spin(4),\myl{b4}
with $h+j=1$ and $h-j=k,$ and $k^2\ \rlap{$\equiv$}/\ 1 {\rm\ mod\ } 7,$
is in fact exotic, i.e., homeomorphic to $S^7$, but not diffeomorphic to
it.  Clearly, the constructive part is  fairly easy, but the proof of the
exotic nature of the resulting sphere is more involved, drawing from
several important results in differential topology including the Thom
bordism result, cohomology theory, Pontrjagin classes, etc.  Later,
Kervaire and Milnor\myc{KM} and others\myc{gromoll} expanded on these
results, leading to a good understanding of the class of exotic spheres
in dimensions seven and greater.\par
For the purposes of this paper, the importance of Milnor's
discovery is that it provides an explicit example of a
topologically simple space, $S^7$, for which the topology does
not determine the smoothness.  Furthermore, its construction is
explicit and easily understood.  From the physics viewpoint, its
most important application may be in terms of some ``exotic
Yang-Mills'' models as discussed below.\par
Unfortunately, the path to $\Rx$ is much less easy to
describe than are the Milnor spheres. The techniques
required span a variety of mathematical disciplines and the
following survey is necessarily brief.
The reader is referred to mathematical reviews for
more details\myc{fuff}.\par
The first step involves a topological tool important in
classifications of four manifolds.
The {\it intersection form} of a compact oriented
manifold without boundary,
is obtained by the Poincare duality pairing of
homology classes in $H_{n-k}$ and $H_k$ can be simply represented in
dimension $n=4=2+2$ by a symmetric square matrix of determinant $\pm 1$.
This form basically reflects the way in which pairs of oriented two-
dimensional closed surfaces fill out the full (oriented) four-space at
their intersection points.  Physicists are perhaps more familiar with
deRham cohomology involving exterior forms for which this intersection
pairing is the volume integral of the exterior product of a pair of
closed two-forms representing the individual cohomology classes, which
again makes sense only in dimension four.  Unfortunately, deRham
cohomology necessarily involves real coefficients and is thus too coarse
for our applications, which need integral homology theory.  At any rate,
this integral intersection
 form, $\omega$,
plays a central role in classifying compact four manifolds.  Whitehead
used it to prove that one-connected closed 4-manifolds are determined up
to homotopy type by the isomorphism class of $\omega.$  Later,
Freedman\myc{fr} proved that $\omega$ together with the Kirby-Siebenmann
invariant classifies simply-connected closed 4-manifolds up to
homeomorphism.  For our purposes, the important result was that there
exists a topological four manifold,
 $\vert E_8\vert,$ having intersection form
$\omega=E_8,$ the Cartan matrix for the exceptional lie algebra of the
same name.  \par
As it stands, Freedman's work is in the topological category,
and does not address smoothness questions.  The theorem of Rohlin\myc{R}
states that the signature of a closed connected oriented
smooth 4-manifold must be divisible by 16,
so that $\vert E_8\vert$ cannot exist
as a {\em smooth} manifold since its signature is 8.
    Next, Donaldson's
theorem\myc{D} provides the crucial (for our purposes) generalization of
this result to establish that
$\vert E_8\oplus E_8\vert$ cannot be endowed with any smooth
structure,
 even though its signature
is 16. The work of Donaldson is based on the moduli space of solutions to
the $SU(2)$ Yang-Mills equations on a four-manifold, which first occur in
physics literature.\par
Having established some algebraic machinery, the next step involves an
algebraic variety, the Kummer surface, $K$, a real four-dimensional
smooth manifold in $CP^3.$  It is known that
\beq
K=\vert -E_8\oplus -E_8 \oplus 3\pmatrix{ 0 & 1 \cr 1 & 0}\vert.\myl{k1}
The last part of this intersection form is easily seen to be realizable
by $3(S^2\times S^2)$, which is smooth.  Thus, Donaldson's theorem
implies that it is impossible to do smooth surgery on $K$ in just such a
way as to excise the smooth $3(S^2\times S^2)$, leaving a smooth
(reversing orientation) $\vert E_8\oplus E_8\vert$. In the following, we
refer to these two parts as $V_1$(smoothable) and $V_2$(not smoothable)
respectively, so smooth $K=V_1\cup V_2.$ In investigating the failure of
this smooth surgery Freedman found the first fake \mRx.  Using a
topological $S^3$ to separate $V_1$ from $V_2$, Donaldson's result showed
that this $S^3$ cannot be smoothly embedded, since otherwise $V_2$ would
have a smooth structure.  However, by further surgery, it is found that
this dividing $S^3$  is also topologically embedded in a topological
\mR\
and actually includes a compact set in its interior. Thus we have\par
\state{Existence of exotic \mRx}
{This topological \mR\ contains a compact
set which cannot be contained in any smoothly embedded $S^3.$  This
surprising result then implies that this manifold is indeed an \mRx\
since
in any diffeomorphic image of
\mR\ every compact set is included in the
interior of a smooth sphere.}\par
\section{Some Properties of \mRx's}\myll{sp}\par
After this basic existence theorem, many developments have
occurred, some of which are
summarized in the book by Kirby\myc{surv}.  Unfortunately, none of the
uncountable infinity of \mRx's
has been presented in explicit atlas of
charts form, so most of the properties can only be described indirectly,
through existence or non-existence type of theorems.
In this paper we restrict the discussion to those topics of
possible
physical applications.
\par
The defining feature of the original \mRx\ as
discussed above can be summarized in coordinate form as
follows: \mRx\ is the {\it topological} product ${\bf R^1\times
 R^1\times R^1\times R^1}$.  Thus, it can be described
 {\it topologically} as the set ${\Rx}=\{p^\alpha\},
 \alpha=1...4.$  However,
these coordinates cannot be globally smooth, since otherwise the
product would be smooth and the manifold diffeomorphic to
standard \mR.  Or, in terms of the property used in the
discovery, if $\{p^\alpha\}$ were globally smooth, since every
compact set is contained in some
$\vert{\bf p}\vert^2\equiv\sum (p^\alpha)^2=R^2$ for
sufficiently large $R$, then every compact set would be contained
in some smoothly embedded $S^3,$
contrary to the existence statement
above.  In fact, this absence of ``sufficiently large'' smooth
three-spheres is a strange and defining characteristic of \mRx's.
 It shows, among other things that the exoticness
necessarily extends to infinity.  Nevertheless, the possibility
of confining the exoticness to a time-like tube is shown
in Theorem 2 below.
\par
Even though the global topological coordinates cannot be globally
smooth, it can be shown that in some diffeomorphic copy they can
be made smooth locally.  Thus\par
 \begin{theorem}
   There exists a smooth copy of each
\mRx\ for which the
global $C\sp 0$ coordinates are smooth in some neighborhood.  That is,
there exists a smooth copy,
${\Rx}=\{(p^\alpha)\}$,
for which $p^\alpha
\in      C\sp \infty$ for $\vert {\bf p}\vert<\epsilon.$
\end{theorem}
\par
The proof of this involves
the Annulus Theorem\myc{annulus}
to $C^0$ tie some local smooth coordinates to the
global $C^0\ \{p^\alpha\}$ across some annulus.\par
What this gives is a local smooth coordinate patch, on which standard
differential geometry can be done, but which cannot be extended
indefinitely.
The obstruction should be physically interesting.  Why cannot
certain local fields be continued globally in the absence of any
topological obstructions?\par
Theorem 1 leads naturally to the following construction.
By puncturing \mRx\ in the neighborhood where the topological
coordinates are smooth,
 we get a ``semi-exotic'' cylinder, i.e,
$\Rx-\{0\}\simeq {\bf R\sp 1}\xt S\sp 3,$  where $\xt$ means topological
but not smooth product.  By ``semi-exotic'' we mean that the product is
actually smooth for a semi-infinite extent of the first coordinate.
This might be a very interesting cosmological model for physics, which
after the big bang is ${\bf R}\sp 1 \times {\bf S\sp 3}.$
 Here we would
run into an obstruction to continuing the smooth product structure at
some finite time (first coordinate) for some unknown,
 but potentially very interesting, reason.\par
An
even more interesting possibility to consider would involve
localizing the ``fakeness'' in some sense.  One version that
comes to mind would happen if we could smoothly glue two such
semi-exotic cylinders at their exotic ends.  Of course a second gluing
at their smooth ends would then give an exotic smoothness on the
topological product,  $S\sp 1 \times S\sp 3$. The existence of such
an   $S\sp 1\xt S\sp 3$ is apparently not known.\par
However, some results on ``localization of exoticness'' can be
obtained.
In fact, the main result of another
 paper\myc{blx} can be summarized informally: \par
\begin{result}
There exists exotic smooth manifolds with ${\bf R^4}$ topology which are
standard at spatial infinity, so that the exoticness can be regarded as
spatially confined.\end{result}\par
A more precise statement of this result is provided in Theorem 2
below.
The resulting manifold structures have the property
 that everything looks normal at space-like
 infinity but the standard structure cannot be continued all the way
in to spatial origin. \par
This result could have great significance in all fields of physics, not
just relativity. Some model of space-time underlies every field of
physics.  It has now been proven that we cannot infer that space  is
necessarily smoothly standard from investigating what happens at space-
like infinity, even for topologically trivial
${\bf R^4}$. It seems very clear that this is potentially very important
to all of physics since it implies that there is another possible
obstruction,  in addition to material sources and topological ones, to
continuing external vacuum solutions for any field equations from
infinity to the origin.  Of course, in the absence of any explicit
coordinate patch presentation, no example can be displayed. However,
this\  leads naturally to a conjecture, informally stated: \par
\begin{conj}
This localized exoticness can act as a source for some externally regular
field, just as matter or a wormhole can.
\end{conj}\par
The full exploration of this conjecture will require more
detailed
knowledge of the global metric structure than is available at present.
The notions of domains of dependence, Cauchy surfaces, etc., necessary
for such studies cannot be fully explored with present differential
geometric information on exotic manifolds.  However, a beginning can be
made with certain general existence results as established and discussed
below.
\par
We now state and sketch the proof of\par
\begin{theorem}
There exists smooth manifolds which are homeomorphic but not
diffeomorphic to ${\bf R^4}$ and for which the global topological
coordinates $(t,x,y,z)$ are smooth for
$x^2+y^2+z^2\ge \epsilon^2>0,$ but not
globally.  Smooth metrics exists for which the boundary of this region is
timelike, so that the exoticness is spatially confined.
\end{theorem}\par
To arrive at this result we make use of techniques
developed by Gompf\cite{G2} which lead to
 the construction of a large
 topological variety of exotic four-manifolds, some of which would
appear to have considerable potential for physics.
 Gompf's ``end-sum'' process  provides a
straightforward technique for constructing an exotic version, $M$, of any
non-compact four-manifold whose standard version, $M_0$, can be
smoothly embedded
in standard ${\bf R^4}$.  Recall that we want to construct $M$\ which is
homeomorphic
to $M_0$, but not diffeomorphic to it.  First construct a tubular
neighborhood, $T_0$, of a half ray in $M_0$. $T_0$ is thus standard ${\bf
R^4}=
[0,\infty)\times {\bf R^3}$.  Now consider a diffeomorphism, $\phi_0$ of
$T_0$
onto
$N_0=[0,1/2)\times{\bf R^3}$ which is the identity on the ${\bf R^3}$
fibers.  Do
the same thing for some exotic ${\bf R^4_\Theta}$\ with the important
proviso that it
{\em cannot} be smoothly embedded in standard ${\bf R^4}$.  Such
manifolds are known
in infinite abundance \cite{G2}.  Then construct a similar
tubular
neighborhood for this ${\bf R^4_\Theta}$, $T_1$, with diffeomorphism,
$\phi_1$, taking
it onto $N_1=[1,1/2)\times{\bf R^3}$.  The desired exotic $M$\ is then
obtained by forming the identification manifold structure
\begin{equation}
M=M_0\cup_{\phi_0}([0,1]\times{\bf R^3})\cup_{\phi_1}\bf
R^4_\Theta\label{es1}\end{equation}
The techniques of forming tubular manifolds and defining
identification manifolds can be found in standard differential
topology texts, such as \cite{bj} or \cite{H}.  \par
Informally, what is being
done is that the tubular neighborhoods are being smoothly glued across
their ``ends'', each ${\bf R^3}$.  The proof that the resulting $M$\ is
indeed  exotic is then easy:  $M$\ contains ${\bf R^4_\Theta}$\ as a
smooth sub-manifold.
If $M$\ were diffeomorphic to $M_0$ then $M$,  and thus ${\bf
R^4_\Theta}$, could be
smoothly embedded in standard ${\bf R^4}$, contradicting the assumption
on
${\bf R^4_\Theta}$.  Finally, it is clear that the constructed $M$\ is
indeed
homeomorphic to the original $M_0$ since all that has been done
topologically is the extension of $T_0$.
\par
In order to relate these constructions to possible physical
applications, and to complete the proof,
let us now introduce $(t,x,y,z)$ as the global
topological coordinates.
Let the tubular
neighborhoods used in the end sum techniques be generated by
the continuous (but not globally smooth) $t$-curves,
and that the standard \mR corresponds to $t<0.$
  Then
``stuffing'' the upper \mRx into the tube  results in a manifold
which we label $M_3,$ having the property that $(t,x,y,z)$ are
smooth for $t<0$ and for $x^2+y^2+z^2>\epsilon^2,$ all $t$ for
some positive $\epsilon$.  An obvious doubling of this property
leads to $M_4$ for which $(t,x,y,z)$ are smooth for all
 $x^2+y^2+z^2>\epsilon^2,$ for all $t.$
The smoothness properties of the $M_4$ can also be stated
in terms of products.
Global $C^0$ coordinates, $(t,x,y,z)$, are
smooth in the exterior region
 $[\epsilon,\infty){\bf\times S^2\times R^1}$, while the closure
of
the complement of this is clearly an exotic ${\bf B^3\times_\Theta
R^1}$.\footnote{Here the ``exotic'' can be understood as referring to the
product which is continuous but cannot be smooth.  See the discussion
around Lemma 2 below.}  Since the exterior component is standard, a wide
variety, including flat, of Lorentz metrics can be imposed.  Picking only
those for which $\partial/\partial t$ is timelike in this region provides
a natural sense in which the world-tube confining the exotic part is
``spatially localized.''  The smooth continuation of such a metric to the
full metric is then  guaranteed by Lemma 1 below
and the discussion following
it below.  This completes the proof.\par
 \section{Some Geometry on \mRx's}\myll{sg}\par
Some very basic, if sketchy,
information about differential geometry on \mRx\
can been
obtained. For example,\par
\begin{theorem}
  There can be no  geodesically complete
metric (of any signature) with non-positive sectional curvature on
\mRx.
\end{theorem}
\par
Proof:  If there were such a metric, the Hadamard-Cartan theorem could
be used to show that the exponential map would provide   a diffeomorphism
of
the tangent space at a point onto \mRx.
   In particular, there can be no flat geodesically complete metric. For
more discussion on exotic geometry, see \myc{Reinhart}.
Natural questions then arise concerning the nature of the obstructions
to continuing the solutions to the differential equations expressing
flatness in the natural exponential coordinates.  In physics,
obstructions to continuation of solutions are often of considerable
significance, e.g., wormhole sources.  However, up to now, such
obstructions generally have been  a result of
either  topology, (incompleteness caused by excision),
or some sort of curvature singularity.  Neither of these is present here.
This
problem is particularly interesting for those
 \mRx 's which cannot be
smoothly embedded in standard
\mR, which thus cannot be geodesically
completed with a flat metric.  \par
Consider now what can be said about the continuation of a Lorentz
signature metric from some local chart.  It turns out
that any Lorentzian metric on a closed
submanifold, $A$, some smooth continuation to all of $M$
 under certain conditions.
  For example, we have
\begin{lemmas}
 If $M$ is any smooth connected 4-manifold and $A$ is a closed
submanifold for which
$H^4(M,A; {\bf Z})=0,$ then any smooth
time-orientable Lorentz signature metric defined
over $A$
can be smoothly continued to all of $M.$
\end{lemmas}

Proof: This is basically a question of the continuation of cross
sections on fiber bundles.  Standard obstruction theory is usually
done in the continuous category, but it has a natural extension to the
smooth class, \cite{st}.
First, we note that any time-orientable
 Lorentz metric is decomposable into a Riemannian one, $g$,
plus a non-zero time-like vector field, $v$.
The continuation of $g$ follows from
the fact that the fiber, $Y_S$, of non-degenerate symmetric four by four
matrices is $q$-connected for all $q$.  From standard obstruction theory,
this implies that $g$ can be continued from $A$ to all
of $M$
without any topological restrictions. On the other hand, the fiber of
non-zero vector fields is the three-sphere which is $q$-connected for
all $q<3,$ but certainly not 3-connected ($\pi_3(S^3)={\bf Z}$).  Again
from standard results,
\cite{st}, any obstruction to a
continuation of $v$ from $A$ to all of $M$
is an element of $H^4(M,A;{\bf Z})$.  Thus, the vanishing of
this group is a sufficient condition for the continuation of $v$,
establishing the Lemma.
\par
What is missing from this result, of course, is that the continued metric
satisfy the vacuum Einstein equations and that it be complete in
an appropriate
Lorentzian sense.  Of course, any smooth Lorentzian metric satisfies
the Einstein equation for some stress-energy tensor, but this tensor
must be shown to be physically acceptable.
 Unfortunately, these issues cannot be resolved
without more explicit information on the global exotic structure than
is presently available.
However, we can be more specific about the conditions under which
some smooth Lorentzian metric can be globally continued on an
exotic manifold from some local coordinate presentation.
\par
In the applications in this paper, $M$ is non-compact, so
$H^4(M;{\bf Z})=0.$ Using the exact cohomology sequence generated by
the
inclusion $A\rightarrow M,$ we can develop\myc{blx} several
easily satisfied
sufficient conditions on $A$ to meet the conditions of Lemma 1.
 One easy one is to
require $H^3(A;{\bf Z})=0.$  Another would be to establish that
the map, $H^3(M;{\bf Z})\rightarrow H^3(A;{\bf Z})$ is an epimorphism.
For  example, if $A$ is simply a closed miniature
version of ${\bf R^2\times S^2}$ itself, i.e.,
$A={\bf D^2\times S^2}$, then $H^3(A;{\bf Z})=0$ so
the continuation of a smooth Lorentzian metric is ensured.
The spaces
 ${\bf R^2\times_\Theta S^2}$, each have the topology of the
Kruskal presentation
of the Schwarzschild metric.  Using the standard Kruskal notation
$\{(u,v,\omega); u^2-v^2<1, \omega\in S^2\}$ constitute global
{\em topological} coordinates.  From Theorem 1, these can be
smooth over the closure of some open set, say $A$,
homeomorphic ${\bf D^2\times S^2}$,
but {\em $(u,v)$ cannot be continued
as smooth functions over the entire range: $u^2-v^2<1$.}
Over $A$ then we can solve the vacuum
Einstein equations as usual to get the Kruskal form.
{}From Lemma 1, {\it some} smooth
 metric can be continued from this over
the entire manifold.  However, whatever it is, it cannot be the
standard Kruskal one.  The obstruction to continuation of the
metric occurs
not for any reasons associated
with the development of
singularities in the coordinate expression of the metric, or for any
topological reasons, but simply because {\em the coordinates,
$(u,v,\omega)$, cannot be continued smoothly beyond some proper
subset, $A$,
of the full manifold}.
This  establishes
\begin{theorem} On some smooth manifolds
which are topologically
${\bf R^2\times S^2}$, the standard Kruskal metric
cannot be smoothly continued over the full range, $u^2-
v^2<1.$
\end{theorem}\par

An interesting variation occurs when $A$ contains a
trapped
surface, so a singularity will inevitably develop from well-known
theorems.  However, if $A$ does not contain a trapped surface
what will happen is
not known.
\par

Another way to study this metric is in terms of
 the original Schwarzschild
$(r,t)$ coordinates for $M_4$.    For this model the
coordinates
$(t,r,\omega)$ are smooth for all of the closed sub-manifold $A$ defined
by
 $r\ge \epsilon>2M$ but cannot be continued
as smooth over the entire $M$ or over any diffeomorphic (physically
equivalent) copy.  In this case $A$ is topologically
$[\epsilon,\infty)\times{\bf S^2\times R^1}$,
 so again $H^3(A;{\bf
Z})=0$
and the conditions of lemma 1 are met.  Hence there is some
smooth continuation of any exterior Lorentzian metric in $A$, in
particular,
 the Schwarzschild metric, over the full ${\bf R^4_\Theta}$.  Whatever
this metric
is, it cannot be Schwarzschild since the manifolds are not diffeomorphic.
  An interesting feature of this model is that the manifold is
``asymptotically'' standard in spite of the well known fact that
exotic manifolds are badly behaved ``at infinity''.  However, we note
that this model is asymptotically standard only as
$r\rightarrow\infty,$ but certainly not as $t\rightarrow\infty.$ \par
These models are clearly
highly suggestive for investigation of alternative continuation of
exterior solutions into the tube near $r=0.$  We often
first discover an {\em exterior}, vacuum solution, and look to
continue
it back to some source.  This is a standard problem.  In the stationary
case, we typically have a local, exterior solution to an elliptic
problem, and try to continue it into the origin but find we
can't as
a vacuum solution unless we have a topology change (e.g., a
wormhole),
or unless we add a matter source, changing the  equation.
Now, we are led to consider a third
alternative, can exotic smoothness serve as a source for some
exterior metric?\par
Of course, the discussion of stationary solutions involves the idea of
time foliations, which cannot exist globally for these exotic
manifolds, at least not into standard factors.
In fact,
\begin{lemmas}
${\bf R^4_\Theta}$\ cannot be written as a smooth product,
${\bf R^1\times_{smooth} R^3}$. Similarly
 ${\bf R^2\times_\Theta S^2}$\ cannot be written as ${\bf
R^1\times_{smooth}(
R^1\times S^2)}$.
\end{lemmas}
Clearly, if either factor decomposition were smooth, the original
manifold would be standard, since the factors are necessarily
standard from known lower dimensional results, establishing
the lemma. It is not
now possible to establish the
 more general result for which the second factor is simply some
smooth three manifold without restriction.\par
Of course, the lack of
a global time foliation of these manifolds means that such models are
 inconsistent with canonical approach to gravity,
quantum theory, etc.  However, it is worth noting that all
experiments yield only local data, so we have no {\em a priori} basis
for excluding such manifolds.\par
These discussions lead naturally to a consideration of what can be
said about Cauchy problems.
Consider then
the manifold, $M_3,$ for which
   the global $(t,x,y,z)$
coordinates are smooth for all $t<0$ but not globally. Now consider,
the Cauchy problem $R_{\alpha\beta}=0$, with flat initial
data on $t=-1$. This is guaranteed to have the complete flat metric
as solution in the standard, ${\bf R^4}$\ case.
However, because of Theorem 3, this cannot be true for $M_3.$
What must go wrong in the exotic case, of course, is
that $t=-1$ is no longer a Cauchy surface.  However, Lemma 1 can again
be applied here to guarantee the continuation of {\em some} Lorentzian
metric over the full manifold since here $A=(-\infty,-1]\times{\bf R^3}$
so clearly $H^3(A;{\bf Z})=0$.  \par
Finally,  consider a cosmological
model based on ${\bf R^1\times_\Theta S^3}$ .  In this
case, we can start with
a standard cosmological metric for some time, so here
$A=(-\infty,1]\times{\bf S^3}$.  Clearly,  $H^3(A;{\bf Z})$ does
not vanish in this case, but it can be shown that the inclusion
induced map
$H^3(M;{\bf Z})\rightarrow H^3(A;{\bf Z})$ is onto, so the conditions
of Lemma 1 are met. Thus  some smooth Lorentzian continuation will indeed
exist, leading to some exotic cosmology on ${\bf R^1\times S^3}$.
  \par

\section{Conclusion and Conjectures}\myll{conc}\par
What are the
possible physical implications of the existence of the exotic
spaces?  First, consider the \msx,
which can be explicitly constructed.  Perhaps they could be considered
as possible models for  exotic Yang-Mills theory. Some
\msx\ are
 $SU(2)$ bundles, but not  principle ones, since their groups must
be $Spin(4)$.  This would contrast with standard Yang-Mills
structure\myc{yama} in which the total space is $S^7$ regarded as a
principle $SU(2)$ bundle.  Next,
\msx\ can be used as  toy space-time
models,
serving as the base manifolds for various geometric and other field
theories.  Various questions of physical interest can then be asked
on these models and the answers compared to those obtained from standard
$S^7$.  For example, the non-existence of a constant curvature metric
on
\msx\ has already been thoroughly explored\myc{gromoll}.  The
analysis
of such differential geometric problems on
\msx\ as compared to $S^7$
 should give some indication of the type of results
that could come from physics on
\mRx\ as compared to that on standard
${\bf R^4}.$
\par
There are also many obvious questions concerning the
physical implications of doing general relativity on \mRx.
The entire problem of developing a manifold from a coordinate
patch piece on which an Einstein metric is known, still has many
unanswered aspects even in standard smoothness.
Recall for example the evolution of our understanding of
the
appropriate manifold to support the (vacuum)
 Schwarzschild metric. Originally,
the solution was expressed using $(t,r,\theta,\phi)$ coordinates
as differentiable outside of the usual ``coordinate singularities''
well known for spherical coordinates.  However, the Schwarzschild
metric form itself in these coordinates exhibits another singularity
on $r=2m$, sometimes referred to as the ``Schwarzschild singularity.''
Later work, culminating in the Kruskal representation, showed that
the Schwarzschild singularity could be regarded as merely another
coordinate one  in the same sense as is the z-axis for $(r,\theta,\phi)$.
This example helps to illustrate that
 in general relativity our understanding of the
physical significance of a particular metric often undergoes an evolution
as various coordinate representations are chosen.  In this process,
the topology and differentiable structure of the underlying manifold
may well change.  In other words, as a practical matter, the study of the
completion of a locally given metric often involves the construction of
the global manifold structure in the process.  Could any conceivable
local Einstein metric lead to  an
\mRx\ by such a process?
\par
Of course, local coordinate patch behavior is of great importance
to physics, so another set of physically interesting questions would
relate to the coordinate patch study of
\mRx.   This may be
too difficult of a task for present mathematical technology, but  some
questions may be reasonable.
For example, can some \mRx's be covered by only a finite
number of coordinate patches?  If so, what is the minimum number?
  What are the intersection
properties of the coordinate patch set which makes it non-standard?\par
A directly physical set of questions to be considered would stem
from attempts to embed known solutions to the Einstein equations in
\mRx,
then asking what sort of obstruction intervenes
to prevent their indefinite, complete, continuation, in this space.
Examples of non-topological obstructions to the continuation
of a wide class of Einstein metrics was discussed in
\mysref{sg}.  What is the physical
significance
of these obstructions?  \par
Finally, it is indeed true that the existence of
\mRx's does not in any
way change the {\em local} physics of general relativity or any other
field theory.  However, it has long been known that global questions can
have profound effects on a physical theory.  Until recently, physicists
have thought of global matters almost exclusively as being of purely {\em
topological} significance, whereas we now know that at least in the
physically important case of ${\bf R^4},$ there are very exciting global
questions related to differentiability structures, the way in which local
physics is patched together smoothly to make it global.  Certainly, the
\mRx's
are essentially just ``other'' manifolds.  However, there are an
infinity of them which have never been remotely considered in the
physical context of classical space-time physics on Einstein's original
model,
\mR.  It would be surprising indeed if none of these had any
conceivable physical significance.\par
 I am very grateful to Duane Randall and Robert Gompf for their
invaluable assistance in this work.
\par

\end{document}